\documentclass[twocolumn,showpacs,preprintnumbers,amsmath,amssymb,prl,aps]{revtex4}

\usepackage{epsfig}
\usepackage{graphicx}
\usepackage{dcolumn}
\usepackage{bm}

\begin{document}

\title{Electronic structure near quantum critical point}

\author{Ganesh Adhikary, R. Bindu, Sudhir K. Pandey, and Kalobaran Maiti}
\altaffiliation{Corresponding author: kbmaiti@tifr.res.in}

\affiliation{Department of Condensed Matter Physics and Materials'
Science, Tata Institute of Fundamental Research, Homi Bhabha Road,
Colaba, Mumbai - 400 005, INDIA.}

\date{\today}

\begin{abstract}
We studied the evolution of the electronic structure across the
quantum critical point in V doped Cr employing high resolution
photoemission spectroscopy. Experimental results exhibit signatures
of pseudogap and orbital Kondo resonance peak at low temperatures
for all the compositions studied suggesting a scenario of spin
density wave quantum criticality corresponding to orbital Kondo
effect. The pseudogap and the Kondo peak gradually reduces with V
doping but remains finite at the quantum critical point indicating
their relevance in widely discussed quantum phases in correlated
electron systems. The spectral lineshape near the Fermi level
exhibit $(\epsilon_F-\epsilon)^{0.25}$ dependence evidencing
deviation from Fermi liquid behavior.
\end{abstract}


\maketitle

Quantum phase transition \cite{sachdev}, the quantum fluctuations
induced phase transition at zero temperature, has attracted a lot of
interests due to its implication in various interesting electronic
properties such as Kondo effect, heavy fermion behavior,
superconductivity {\it etc}. Usually, a non-thermal control
parameter such as pressure, magnetic field or composition is varied
to achieve the quantum critical point \cite{friedemann}. Different
studies on wide range of materials in the vicinity of quantum
critical point shows the breakdown of Landau's Fermi liquid theory,
particle-hole asymmetry etc.\cite{Anderson,RaviEPL} Probing
microscopic origin of such novel phenomena is often hindered by
complexity of the materials involved. Recently, it was shown that
quantum phase transition can be achieved \cite{yeh} in elemental Cr
by V substitution. Cr is an incommensurate spin density wave-type
antiferromagnet below a transition temperature, $T_N$ = 311 K
\cite{Fawcett}. $T_N$ gradually reduces with V substitution and
becomes zero around 3.5\% V concentration, which is defined as
quantum critical point (QCP) due to the finding of non Fermi liquid
behavior in the electronic properties. These experimental results
were interpreted considering the pinning of the Fermi level,
$\epsilon_F$ at the minimum of the electronic density of states
(DOS) \cite{bandstructure} and importance of the antiferromagnetic
spin fluctuations \cite{spinfluc1,spinfluc2}.

Enormous efforts (both experimental and theoretical) have been put
forward to understand the underlying physics exhibiting varied
scenario even in the parent compound, Cr such as suggestion of
multiple gaps in an infrared reflectivity study \cite{boekelheide},
presence of pseudogap in a theoretical study \cite{pepin}, orbital
Kondo resonance in a scanning tunneling spectroscopic study
\cite{Yu,Yuprb} {\it etc}. Conflicting scenarios such as Kondo
interactions, electron-phonon coupling, Shockley-type surface state
formation, etc. have been proposed to capture the complex electronic
properties of Cr
\cite{wiesendanger,donath,CrAPL,klebanoff,gewinner}. Photoemission
studies indicated two back-folded band below $T_N$ in the
incommensurate state and one back-folded band in the commensurate
phase \cite{review1,review2,review3,schafer}. The incommensurability
in the antiferromagnetic phase was found to increase with increasing
V content \cite{krupin}. Signature of non-Fermi liquid behavior was
not detected in this study with an energy resolution of 30 meV
\cite{krupin} suggesting necessity of better experimental conditions
to probe the behavior near QCP. Here, we employed high energy
resolution in the photoemission spectroscopy to probe the electronic
structure around quantum criticality in V-doped Cr. The experimental
results exhibit signature of a pseudogap and a Kondo resonance peak
in Cr that persists across the quantum critical point. The evolution
of the spectral DOS at $\epsilon_F$ with V doping is consistent with
the bulk results \cite{yeh}. The lineshape of the low temperature
spectral function exhibit evidence of non-Fermi liquid behavior.

The samples of V-doped Cr were prepared by congruent melting of
stoichiometric amount of Cr and V in a high purity argon atmosphere
using an water cooled arc-furnace and the homogeneity was achieved
by long sintering. The V concentration was determined by energy
dispersive analysis of $x$-rays (EDAX) and found to be close to the
starting compositions. All the samples were characterized by $x$-ray
diffraction (XRD) exhibiting single phase. The magnetic measurements
on Cr and Cr$_{0.98}$V$_{0.02}$ exhibit incommensurate spin density
wave-type antiferromagnetic transition at 311 K and 150 K,
respectively. No magnetic transition was observed in 3.5\% and 5\%
doped compositions consistent with the earlier results \cite{yeh}.
The photoemission experiments were performed on scraped surface in a
spectrometer equipped with SES2002 Gammadata Scienta analyzer,
monochromatic He {\scriptsize I} ($h\nu$ = 21.2 eV), He {\scriptsize
II} ($h\nu$ = 40.8 eV), Al $K\alpha$ ($h\nu$ = 1486.6 eV) photon
sources and an open cycle helium cryostat LT-3M from Advanced
Research Systems. The energy resolutions were fixed to 2.5 meV, 5
meV and 350 meV for He {\scriptsize I}, He {\scriptsize II} and Al
$K\alpha$ measurements, respectively. Experiments were carried out
at a base vacuum better than 3$\times$10$^{-11}$ torr and
temperature down to 10 K.

\begin{figure}
 \vspace{-2ex}
\includegraphics [scale=0.4]{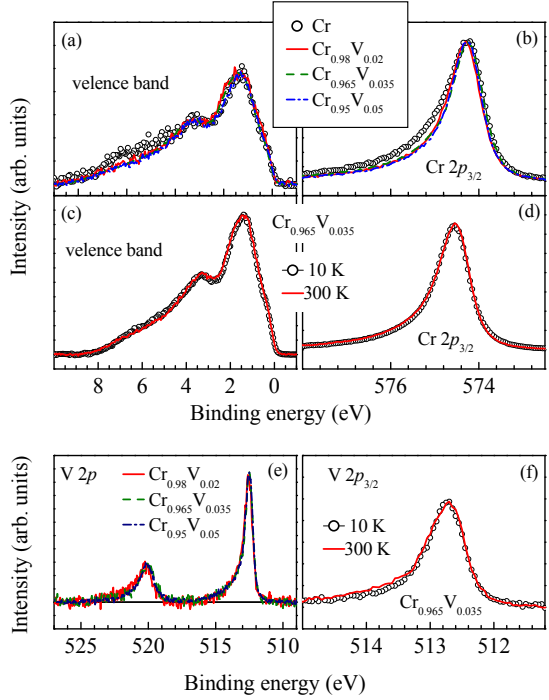}
 \vspace{-12ex}
\caption{(a) Valence band and (b) Cr 2$p_{3/2}$ spectra of
Cr$_{1-x}$V$_x$ at 300 K. (c) Valence band and (d) Cr 2$p_{3/2}$
spectra of Cr$_{0.965}$V$_{0.035}$ at 10 K (open symbols) and 300 K
(line). (e) V 2$p$ spectra of Cr$_{1-x}$V$_x$ at 300 K. (f) V
2$p_{3/2}$ spectra of Cr$_{0.965}$V$_{0.035}$ at 10 K (symbols) and
300 K (line).}
 \vspace{-2ex}
\end{figure}

In Fig. 1(a), we show the room temperature valence band spectra of
various compositions using Al $K\alpha$ radiations, which exhibit
two distinct peaks around 1 eV and 3 eV binding energies due to
$t_{2g}$ and $e_g$ bands \cite{CrAPL}. The spectral lineshape is
almost identical for all the samples studied; this is not unexpected
as the V substitution is small and large resolution broadening in
the $x$-ray spectroscopy. The Cr 2$p_{3/2}$ core-level spectra of
the whole series is shown in Fig. 1(b). The asymmetry in the
2$p_{3/2}$ spectrum of Cr is larger than that of the doped samples.
The asymmetry in the core level spectra of a metal arises due to the
low energy excitations across $\epsilon_F$ associated to the
photo-excitation process \cite{doniach-sunjik}. The decrease in
asymmetry with V doping indicates poorer degree of low energy
excitation in the doped compositions and/or reduction in bandwidth
presumably due to the disorder induced localization of the
electronic states \cite{disorder1,disorder2}. The influence of the
change in temperature on the $x$-ray photoemission spectra has been
investigated for all the samples. We did not observe any change in
the spectral lineshape - a typical case is shown in Fig. 1(c) and
1(d), where we show the valence band and Cr 2$p_{3/2}$ core level
spectra of Cr$_{0.965}$V$_{0.035}$ at 10 K and 300 K suggesting
large energy scale changes to be negligible in these systems.

In Fig. 1(e), we show the V 2$p$ core level spectra exhibiting two
peaks around 512.5 eV and 520 eV binding energies corresponding to V
2$p_{3/2}$ and 2$p_{1/2}$ photoemission - spin-orbit splitting is
found to be about 7.5 eV and the branching ratio
(2$p_{1/2}$/2$p_{3/2}$ intensity ratio) is about 1:2 as expected
from the multiplicity of 2$p$ levels. The 2$p$ spectra are found to
be almost identical in the whole composition range. Change in
temperature does not have significant influence in the spectral
lineshape as evidenced in a typical case of the 2$p_{3/2}$ spectra
of Cr$_{0.965}$Cr$_{0.035}$ in Fig. 1(f).

\begin{figure}
 \vspace{-2ex}
\includegraphics [scale=0.4]{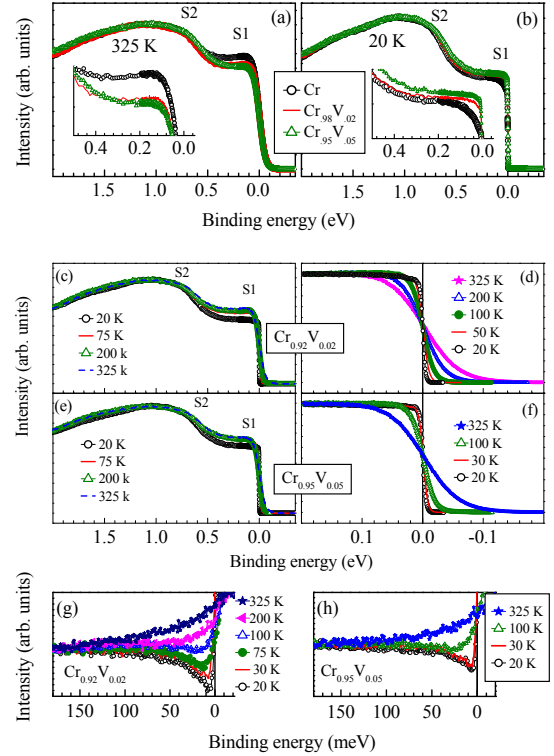}
 \vspace{-4ex}
\caption{The He {\scriptsize I} spectra of Cr,
Cr$_{0.98}$V$_{0.02}$, and Cr$_{0.95}$V$_{0.05}$ at (a) 325 K and
(b) 20 K. The insets show rescaled near $\epsilon_F$ part.
Temperature evolution of the valence band of (c)
Cr$_{0.98}$V$_{0.02}$ and (e) Cr$_{0.95}$V$_{0.05}$. The intensities
near $\epsilon_F$ are shown in an expanded scale for (d)
Cr$_{0.98}$V$_{0.02}$ and (f) Cr$_{0.95}$V$_{0.05}$, respectively.
Spectral DOS of (g) Cr$_{0.98}$V$_{0.02}$ and (h)
Cr$_{0.95}$V$_{0.05}$ obtained by resolution broadened Fermi
function division.}
 \vspace{-2ex}
\end{figure}

In order to investigate the electronic states close to $\epsilon_F$
with better energy resolution, we show the He {\scriptsize I}
valence band spectra for Cr, Cr$_{0.98}$V$_{0.02}$,
Cr$_{0.95}$V$_{0.05}$ at 325 K and 20 K in Fig. 2(a) and 2(b),
respectively. The valence band spectra exhibit four features - the
features around 0.7 eV (S2) and near $\epsilon_F$ (S1) represent the
photoemission signals of the surface electronic structure
\cite{donath,CrAPL,klebanoff,gewinner}. S2 possesses
$d_{xz}d_{yz}$-character and S1 has dominant d$_{z^2}$-character.
The bulk features appear at about 1 eV and 3 eV binding energies
consistent with the $x$-ray photoemission data. From Fig. 2(a) it is
clear that in the paramagnetic state, the intensity of the feature
S1 reduces gradually with the increase in V-content, while all other
features remains almost similar. This is shown in the inset in an
expanded scale. Since, V contains one less electron than Cr, V
substitution in Cr reduces electron counts in the valence band.
Thus, the decrease in intensity near $\epsilon_F$ may be attributed
to such effective hole doping. However, the sharp change in
intensity with 2\% substitution and subsequent weaker change at 5\%
doping seems anomalous.

The 20 K spectra shown in Fig. 2(b) exhibit a different scenario.
While the intensity of S1 is significantly reduced compared to its
intensity in the 325 K spectra in every case, the decrease in the
doped case is much smaller. Interestingly, the change in intensity
near $\epsilon_F$ exhibit trend opposite to that found at 325 K. The
surface contribution at $\epsilon_F$ is known to be negligible in
the low temperature spectra of Cr due to the increase in exchange
interaction strength leading to a shift of the $d_{z^2}$ states
above $\epsilon_F$ \cite{klebanoff}. With the increase in V
substitution, the magnetic ordering temperature reduces drastically
and 5\% doped sample does not show magnetic order. This is clearly
manifested in Fig. 2(c) and 2(e), where we show the He {\scriptsize
I} spectra of Cr$_{0.98}$V$_{0.02}$ and Cr$_{0.95}$V$_{0.05}$ at
different temperatures. In both the compositions, the feature S1
exhibit reduction with temperature, but the reduction is weaker in
Cr$_{0.95}$V$_{0.05}$ that leads to an opposite trend in the low
temperature spectra.

We now turn to the spectral evolution in the vicinity of the Fermi
level with high energy resolution shown in Fig. 2(d) and 2(f) for
Cr$_{0.98}$V$_{0.02}$ and Cr$_{0.95}$V$_{0.05}$ respectively. The
lineshape of the spectra at different temperatures indicate
evolution akin to a simple Fermi liquid system in both the cases. We
have extracted the spectral density of states (SDOS) by dividing the
experimental data by the resolution broadened Fermi Dirac
Distribution Function at corresponding temperature. The resulting
SDOS are shown in Fig. 2(g) and 2(h) for Cr$_{0.98}$V$_{0.02}$ and
Cr$_{0.95}$V$_{0.05}$, respectively. The intensity around
$\epsilon_F$ in the 325 K data increases gradually with the decrease
in binding energy indicating the presence of a peak at or above
$\epsilon_F$. The decrease in temperature leads to a reduction in
the spectral intensity gradually. Signature of a dip around
$\epsilon_F$ appears below the antiferromagnetic transition
temperature, $T_N$ = 150 K in Cr$_{0.98}$V$_{0.02}$. Interestingly,
the SDOS of non-magnetic Cr$_{0.95}$V$_{0.05}$ also exhibit a dip
below 100 K as observed in varieties of other compounds
\cite{lab6,cuprates}. In fact, the magnetic moment is negligible at
this composition although the carrier density and magnetic
susceptibility exhibit a higher value \cite{yeh}. A careful look at
the intensity change indicates that the lowest intensity does not
appear at $\epsilon_F$, instead the spectral function exhibit an
upturn indicating the presence of a peak above $\epsilon_F$ in both
the cases.

\begin{figure}
 \vspace{-2ex}
\includegraphics [scale=0.4]{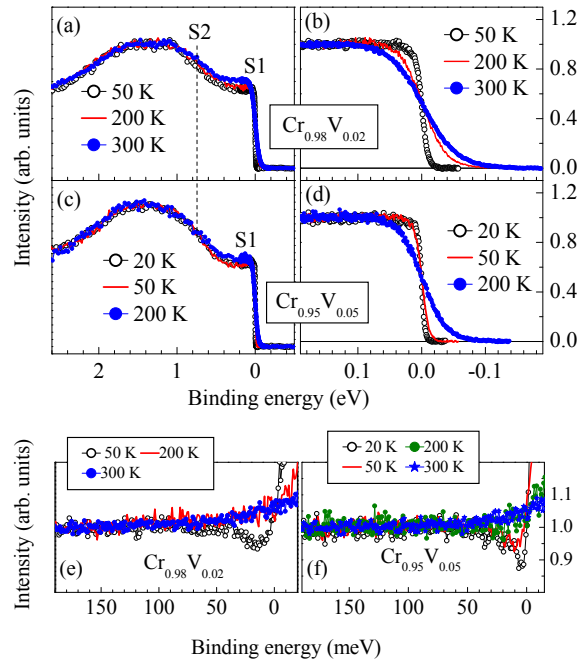}
 \vspace{-12ex}
\caption{Temperature evolution of the He {\scriptsize II} valence
band spectra of Cr$_{0.98}$V$_{0.02}$ in (a) wide energy scale and
(b) near $\epsilon_F$ region, and that of Cr$_{0.95}$V$_{0.05}$ in
(c) wide energy scale and (d) near $\epsilon_F$ region. The spectral
DOS of (e) Cr$_{0.98}$V$_{0.02}$ and (f) Cr$_{0.95}$V$_{0.05}$ at
different temperatures.}
 \vspace{-2ex}
\end{figure}

In Fig. 3, we present the data collected using He {\scriptsize II}
photon energy. Since, valence band of Cr and V-doped Cr primarily
consists of 3$d$ electronic states, the change in photoemission
cross-section due to the change in photon energy will have
negligible influence in the spectral lineshape except a small
enhancement in bulk sensitivity \cite{csvo}. In Fig. 3(a) and 3(c),
we show the valence band spectra at different temperatures for
Cr$_{0.98}V_{0.02}$ and Cr$_{0.95}V_{0.05}$, respectively. As
observed in Fig. 2, the intensity of S1 reduces gradually with
respect to S2 for both the compositions and the rate of decrease is
less significant in the 5\% doped sample. In Fig. 3(b) and 3(d), we
show the spectra very close to $\epsilon_F$ after normalizing them
around 200 meV binding energy. The temperature evolution appears
quite similar to the He {\scriptsize I} spectra. The SDOS obtained
by dividing the experimental data with the resolution broadened
Fermi-Dirac function are shown in Fig. 3(e) and 3(f) for
Cr$_{0.98}V_{0.02}$ and Cr$_{0.95}V_{0.05}$, respectively. In Fig.
3(e), the intensity distribution around $\epsilon_F$ is somewhat
flat. A distinct dip appears at 50 K, which is below the
antiferromagnetic transition temperature. The SDOS of the 5\% doped
sample also exhibit a dip near $\epsilon_F$ as observed in the He
{\scriptsize I} spectra.

\begin{figure}
 \vspace{-2ex}
\includegraphics [scale=0.4]{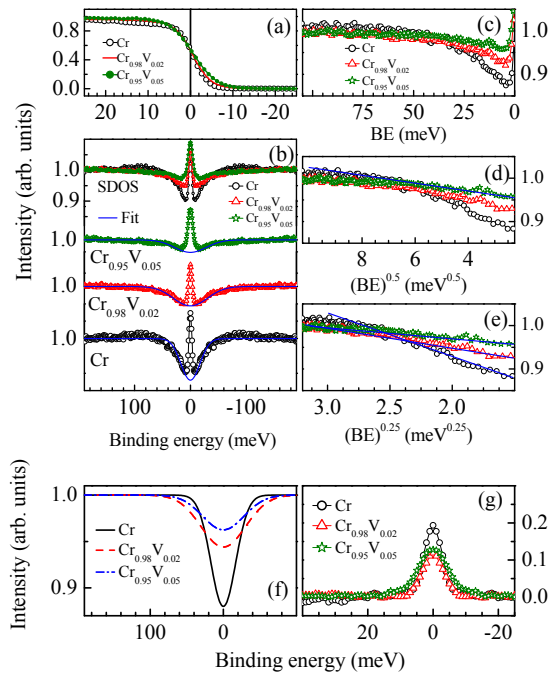}
 \vspace{-12ex}
\caption{(a) He {\scriptsize I} spectra at 20 K of Cr,
Cr$_{0.98}$V$_{0.02}$, and Cr$_{0.95}$V$_{0.05}$. (b) Spectral DOS
at 20 K (symbols) obtained by the symmetrization. The lines are the
polynomial fit away from $\epsilon_F$ representing pseudogap.
Spectral DOS obtained by the resolution broadened Fermi function
division are shown as a function of (c) binding energy (BE), (d)
(BE)$^{0.5}$ and (e) (BE)$^{0.25}$. The lines drawn over the symbols
show linearity of spectral DOS. (f) Comparison of the pseudogaps
obtained in (b). (g) Spectral functions obtained by subtracting the
lines from the spectral DOS shown in (b).}
 \vspace{-2ex}
\end{figure}

From the above results, it is evident that the low temperature
spectral function of all the samples studied exhibit a dip at
$\epsilon_F$ and a peak. While the dip can be attributed to the long
range ordering in the magnetic samples \cite{CaSrRuO}, the presence
of such dip in 5\% doped sample is unusual. It is also important to
investigate the spectral evolution across the quantum critical
point. In Fig. 4, we study the 20 K data of all the compositions in
more detail to understand this phenomena. The raw data, the
corresponding SDOS obtained by symmetrization and SDOS obtained by
the division of the resolution broadened Fermi Dirac function are
shown in Fig. 4(a), 4(b) and 4(c), respectively. Subtle change in
the spectral lineshape is observed in Fig. 4(a). This is most
evident in SDOS shown in Fig. 4(b) and 4(c). Fig. 4(c) exhibits a
large dip just below $\epsilon_F$ in Cr, which gradually reduces
with the increase in V concentration. In addition, a peak like
structure is observed at $\epsilon_F$.

Although the symmetrization smears out asymmetry relative to
$\epsilon_F$, such an exercise is often useful to inspect the SDOS
around $\epsilon_F$ that provides a lower bound of the width of the
pseudogap and/or peak appearing above $\epsilon_F$. Most
importantly, the intensity at $\epsilon_F$ are estimated quite
accurately in this process. Thus, we investigate the symmetrized
SDOS in Fig. 4(b) exhibiting a sharp peak at $\epsilon_F$ in
addition to the dip representing the pseudogap. We have estimated
the pseudogap by fitting the data with a polynomial as shown by
solid lines superimposed over the SDOS. These lines are redrawn
together in Fig. 4(f). The pseudogap gradually reduces with the
increase in V doping. From the bulk measurements, it has been found
that the carrier density and magnetic susceptibility exhibit an
increase at the quantum critical point, where the magnetic moment
and $T_N$ vanishes \cite{yeh}. The reduction of the pseudogap shown
in Fig. 4(f) is consistent with this picture.

The sharp feature at $\epsilon_F$ is delineated by subtracting the
pseudogap lineshape from the SDOS shown in Fig. 4(b) - the resulting
spectra are shown in Fig. 4(g). A sharp peak at $\epsilon_F$ is
evident in the case of Cr due to the orbital Kondo effect
\cite{Yu,Yuprb,CrAPL}. Interestingly, the Kondo resonance peak
appears to survive across the quantum critical point and does not
become stronger in the paramagnetic phase. A typical Doniach phase
diagram, representing competition between Kondo effect and long
range order in a correlated electron system, is somewhat different.
Here, the Kondo peak is strong in the non magnetic Fermi liquid
phase (strong coupled regime) and gradually decreases in intensity
as one moves towards the long range ordered phase. The peak often
vanishes at the quantum critical point, termed as local quantum
criticality \cite{localQCP}. The persistence of Kondo resonance peak
within the long range ordered phase \cite{Ce2MSi3} manifests the
other scenario, namely, spin density wave quantum criticality. The
observation of Kondo peak and long range antiferromagnetic order in
low V doped compositions in the present case resemble a similar
scenario corresponding to orbital Kondo effect.

In order to probe the origin of non-Fermi liquid further, we plot
the SDOS as a function of different exponents of the binding energy,
$BE = (\epsilon_F - \epsilon)$; linear in Fig. 4(c), $(BE)^{0.5}$ in
Fig. 4(d) and $(BE)^{0.25}$ in Fig. 4(e). Clearly, the best
linearity of the lineshape is observed in Fig. 4(e) for all the
cases. If disorder play the dominant role, the exponent is found
usually 0.5 \cite{altshuler,DDdisorder}. In the case of higher
degree of magnetic interactions, electron-magnon coupling etc, the
exponent found to be higher than 0.5, and often it is 1.5
\cite{bairo3}. In the present case, we find much smaller exponent
($\sim$ 0.25), which indicates significant deviation from the Fermi
liquid behavior. To our knowledge, such spectral lineshape is
unusual and clearly needs further study for an understanding.

In summary, the high energy resolution employed in this study
revealed interesting subtle features in the vicinity of quantum
critical point. Experimental data exhibit signature of a Kondo peak
in the spin density wave phase that gradually reduces in intensity
with V doping across QCP. The existence of Kondo peak within
magnetically ordered phase and its persistence at QCP indicates a
scenario of spin density wave quantum criticality in the case of
orbital Kondo effect. In addition, we observe a pseudogap in Cr,
which gradually decreases with increase in V doping. Such a change
leads to an overall increase in the density of states at the Fermi
level consistent with the bulk results. The spectral lineshape in
the vicinity of the Fermi level exhibit
$(\epsilon_F-\epsilon)^{0.25}$ dependence - an evidence of the
deviation from Fermi liquid behavior.

\end{document}